\documentclass[11 pt,a4paper,usenames,dvipsnames]{article}
\usepackage[utf8]{inputenc}
\usepackage{amsmath}
\usepackage{amsfonts}
\usepackage{amssymb}
\usepackage{pgfplots}
\usepackage{xcolor}

\usepgfplotslibrary{polar}
\usepackage{tikz}
\usepackage{subcaption}
\usepgfplotslibrary{fillbetween}
\usetikzlibrary{patterns}
\usepackage{indentfirst}
\usepackage{amsmath}
\usepackage{pdflscape}

\author{Dr. Jacques Balayla MD, MPH, CIP, FRCSC\footnote{To whom correspondence should be addressed: Dr. Jacques Balayla MD, MPH, CIP, FRCSC. e-mail: jacques.balayla@mcgill.ca. Osler Fellow. Department of Obstetrics and Gynecology. Faculty of Medicine. McGill University, Montreal, Quebec, Canada}}

\title{The SIR-P Model:\\ An Illustration of the Screening Paradox}

\date{}

\begin{document}
\maketitle  
\begin{abstract}

In previous work by this author, the screening paradox - the loss of predictive power of screening tests over time $t$ - was mathematically formalized using Bayesian theory.  Where $J$ is Youden's statistic, $b$ is the specificity of the screening test and $\phi$ is the prevalence of disease, the ratio of positive predictive values at subsequent time $k$, $\rho(\phi_{k})$, over the original $\rho(\phi_{0})$ at $t_0$ is given by:
\\
\
\begin{large}
\begin{equation}
\zeta(\phi_{0},k) = \frac{\rho(\phi_{k})}{\rho(\phi_{0})} =\frac{\phi_k(1-b)+J\phi_0\phi_k}{\phi_0(1-b)+J\phi_0\phi_k}
\end{equation}
\end{large} 
\\
\
Herein, we modify the traditional Kermack-McKendrick SIR Model to include the fluctuation of the positive predictive value $\rho(\phi)$ (PPV) of a screening test over time as a function of the prevalence threshold $\phi_e$. We term this modified model the SIR-P model. Where a = sensitivity, b = specificity, $S$ = number susceptible, $I$ = number infected, $R$ = number recovered/dead, $\beta$ = infectious rate, $\gamma$ = recovery rate, and $N$ is the total number in the population, the predictive value $\rho(\phi,t)$ over time $t$ is given by:
\\
\
\begin{large}
\begin{equation}
\rho(\phi,t) = \frac{a[\frac{\beta IS}{N}-\gamma I]}{ a[\frac{\beta IS}{N}-\gamma I]+(1-b)(1-[\frac{\beta IS}{N}-\gamma I])}
\end{equation}
\end{large}
\\
\
Otherwise stated:
\begin{large}
\begin{equation}
\rho(\phi,t) = \frac{a\frac{dI}{dt}}{ a\frac{dI}{dt}+(1-b)(1-\frac{dI}{dt})}
\end{equation}
\end{large}
\end{abstract} 
\

where $\frac{dI}{dt}$ is the fluctuation of infected individuals over time $t$. 
\newpage

\section{Compartmental models in epidemiology}
The use of modelling in epidemiology provides insight into the understanding of disease dynamics. In particular, compartmental models, which assign disease status that can vary over time to a well-defined cohort, are particularly useful in the study of outbreaks and epidemics. For the proper assessment of disease dynamics in $real$ $time$ - an understanding of the intricacies of the screening process is critical. Unfortunately, concepts such as the proportionality between a tests' predictive value and the prevalence of disease, the screening paradox, and the prevalence threshold are seldom accounted for in epidemiologic models of disease. Towards this end, previous work by this author provided a formalization of the screening paradox, a summary of which is provided below. In order to gain insight into the intricacies of the screening process and how its reliability varies over time, we hereby illustrate and integrate the screening paradox into a basic compartmental SIR model.

\section{The Screening Paradox}
Given the shape of the screening curve (Predictive value as a function of disease prevalence), and the principle of the prevalence threshold, even small changes in the prevalence $\phi$ can have significant changes in the positive predictive value  $\rho(\phi)$. To determine the degree of reduction in the predictive value of the screening test at time $t_k$, we take the ratio of $\rho(\phi)$ at two different times, be it $t_0$, and some later time $t_k$ with a prevalence reduction of $\phi_{0}-k$, where $k<\phi$ is the percentage reduction in prevalence:
\\
\
\begin{large}
\begin{equation}
 \frac{\rho(\phi_{0}-k)}{\rho(\phi_{0})}  = \frac{\left(\phi-k\right)\left[a\phi+\left(1-b\right)\left(1-\phi\right)\right]}{\phi\left[a\left(\phi-k\right)+\left(1-b\right)\left(1+k-\phi\right)\right]}
\end{equation}
\end{large} 
\

Since $\phi_{0}-k$ yields a new, lower prevalence $\phi_{k}$, we can re-write the above equation as:

\begin{large}
\begin{equation}
 \frac{\rho(\phi_{k})}{\rho(\phi_{0})}  = \frac{\phi_{k}\left[a\phi_0+\left(1-b\right)\left(1-\phi_0\right)\right]}{\phi_0\left[a\phi_{k}+\left(1-b\right)(1-\phi_{k})\right]}
\end{equation}
\end{large} 
\

The term $\varepsilon-1 = a+b-1$ has been previously defined in the context of receiver-operating characteristics (ROC) curves, and is termed the Youden's $J$ statistic \cite{youden1950index}. As such, we can re-write the above equation as:
\begin{large}
\begin{equation}
\zeta(\phi_{0},k) = \frac{\rho(\phi_{k})}{\rho(\phi_{0})} =\frac{\phi_k(1-b)+J\phi_0\phi_k}{\phi_0(1-b)+J\phi_0\phi_k} 
\end{equation}
\end{large} 
\
From the above relationship, we infer:
\begin{large}
\begin{equation}
\lim_{k \to 0}\frac{\rho(\phi_{k})}{\rho(\phi_{0})}=\lim_{k \to 0}\frac{\rho(\phi_{0}-k)}{\rho(\phi_{0})}=\lim_{k \to 0}\zeta = 1
\end{equation}
\end{large}

$\zeta(\phi_{0},k)$ may be considered as the predictive value percentage loss as the prevalence decreases from $\phi_0$ to $\phi_k$.
\\
\

Since the positive predictive value of a test is prevalence-dependent, the screening paradox can be explained as follows: if a disease process is screened for and subsequently treated, its prevalence would drop in the population, which as per Bayes' theorem, would make the tests’ predictive value drop in return. Put another way, a very powerful screening test would, by performing and succeeding at the very task it was developed to do, paradoxically reduce its ability to correctly identify individuals with the disease it screens for in the future over some time $t$. Simply stated, the screening paradox affirms that an increase in screening eventually leads to less, or more accurately, lower quality screening as the prevalence drops below the prevalence threshold $\phi_e$. The mechanism by which the screening paradox arises is depicted through the following arrow flow diagram (Figure 1):
\\
\
\begin{center}
\fbox{\begin{minipage}{24em}
Given the presence of a disease amenable to screening:
\\
\

As per the $Wilson-Jungner$ criteria \cite{wilson1968principles}:
\

\begin{center}
1) $\uparrow$ Screening $\rightarrow$ $\uparrow$ Treatment
\
\end{center}
As per the axiom of prevalence \cite{baldessarini1983predictive}:
\

\begin{center}
2) $\uparrow$ Treatment $\rightarrow$ $\downarrow$ Prevalence
\
\end{center}
As per Bayes' Theorem \cite{balayla2020prevalence}:
\

\begin{center}
3) $\downarrow$ Prevalence $\rightarrow$ $\downarrow$ Positive Predictive Value
\
\end{center}
As per the principles of consumer value and utility \cite{winsten1981competition}:
\

\begin{center}
4) $\downarrow$ Positive Predictive Value $\rightarrow$ $\downarrow$ Screening
\\
\

\end{center}
\end{minipage}}
\end{center}
\newpage

\section{Modified SIR: the SIR-P Model}
To better illustrate the ideas above, we can take an infectious disease we shall call "X" for simplicity's sake. The condition need not be infectious in nature, but an infectious agent lends itself well to the application of the concepts herein described. To estimate how the prevalence of disease X changes over time in a community outbreak, we can come up with a theoretical SIR model \cite{weiss2013sir}. An SIR model is an epidemiological model that computes the theoretical number of people infected with a contagious illness in a closed population over time. The name of this class of models derives from the fact that they involve coupled ordinary differential equations relating the number of susceptible people S(t), number of people infected I(t), and number of people who have recovered R(t) over time $t$ \cite{smith2004sir}. One of the simplest SIR models is the Kermack-McKendrick model \cite{wang2016traveling}. The dynamics of an infectious epidemic, such as the case of X, are often much faster than the dynamics of birth and death due to other causes than X, therefore, birth and death are often omitted in simple compartmental models. Otherwise stated - the population remains relatively stable over time $t$ so that the individual parameters can change, but the total number of individual remains stable:

\begin{large}
\begin{equation}
S(t) + I(t) + R(t) = N
\end{equation}
\end{large}

where N is some constant.
\\
\

The SIR system can be expressed by the following set of ordinary differential equations:

\begin{large}
\begin{equation}
\frac{dS}{dt} = -\frac{\beta IS}{N}
\end{equation}
\end{large}

\begin{large}
\begin{equation}
\frac{dI}{dt} = \frac{\beta IS}{N}-\gamma I
\end{equation}
\end{large}

\begin{large}
\begin{equation}
\frac{dI}{dt} = \gamma I
\end{equation}
\end{large}

where $\beta$ is the average infection rate, $\gamma$ is the recovery rate, $S$ is the proportion of susceptible population, $I$ is the proportion of infected, $R$ is the proportion of removed population (either by death or recovery), and $N$ is the sum of the latter three. From the above equations we obtain the basic reproduction number ($R_0$) as a ratio of infection-to-recovery rates \cite{dietz1993estimation}:
\begin{large}
\begin{equation}
R_0 = \frac{\beta}{\gamma}
\end{equation}
\end{large}

The ensuing equation to determine the number of susceptible individuals as a function of time becomes:

\begin{large}
\begin{equation}
S(t)=S(0)e^{-R_{0}(R(t)-R(0))/N}
\end{equation}
\end{large}

where S(0)and R(0) are the initial numbers of,  susceptible and recovered/dead subjects, respectively.  The graphical representation of the above dynamics over some time $t$ is seen in Figure 2:
\begin{figure}[hbt!]
\centering
\includegraphics[scale=0.5]{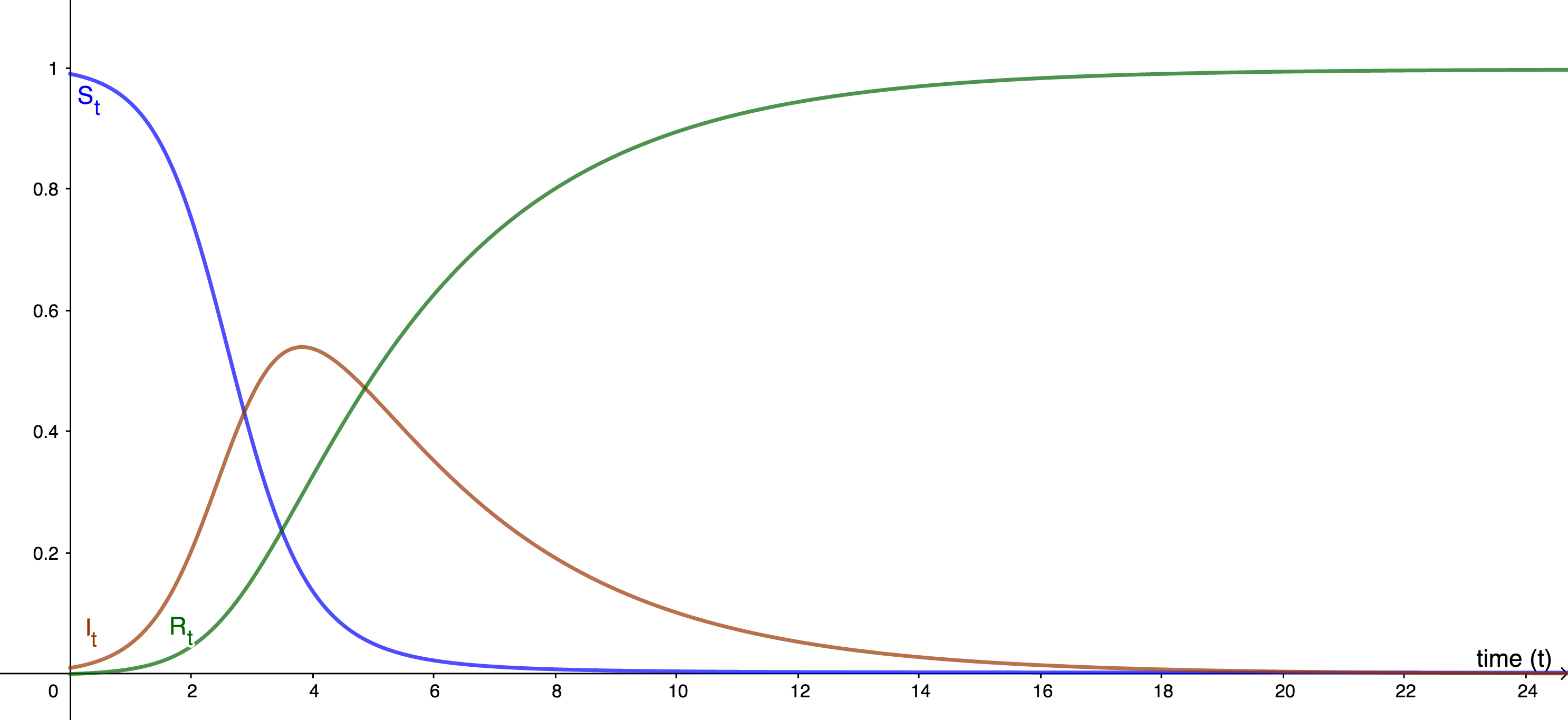}
\\
\

\textbf{Figure 2}. Dynamics of the SIR Model
\end{figure}

For the purpose of the screening paradox, we need only focus on equation 10, the rate of change of active infections $dI$/$dt$, which as a rate reflects the incidence of disease, but as an absolute value in a specific time $t$ yields the prevalence of disease X at that point in time (Figure 3). We can use this value to determine how the PPV fluctuates over time. The differential equation relating the changes in PPV over time $t$ therefore becomes:

\begin{small}
\begin{equation}
\rho(\phi,t) = \frac{a[\frac{\beta IS}{N}-\gamma I]}{ a[\frac{\beta IS}{N}-\gamma I]+(1-b)(1-[\frac{\beta IS}{N}-\gamma I)]}= \frac{a\frac{dI}{dt}}{ a\frac{dI}{dt}+(1-b)(1-\frac{dI}{dt})}
\end{equation}
\end{small}

Towards this end, let us assume that the condition X has a screening test with excellent sensitivity and specificity parameters of 95 and 99 percent, respectively. According to equation 12, this test would have a prevalence threshold of 9.3 percent. Using this threshold, we can illustrate the full SIR model as follows (Figure 3):

\begin{landscape}
\begin{figure}[]
\textbf{Figure 3}. SIR model and PPV
\centering
\includegraphics[scale=0.24]{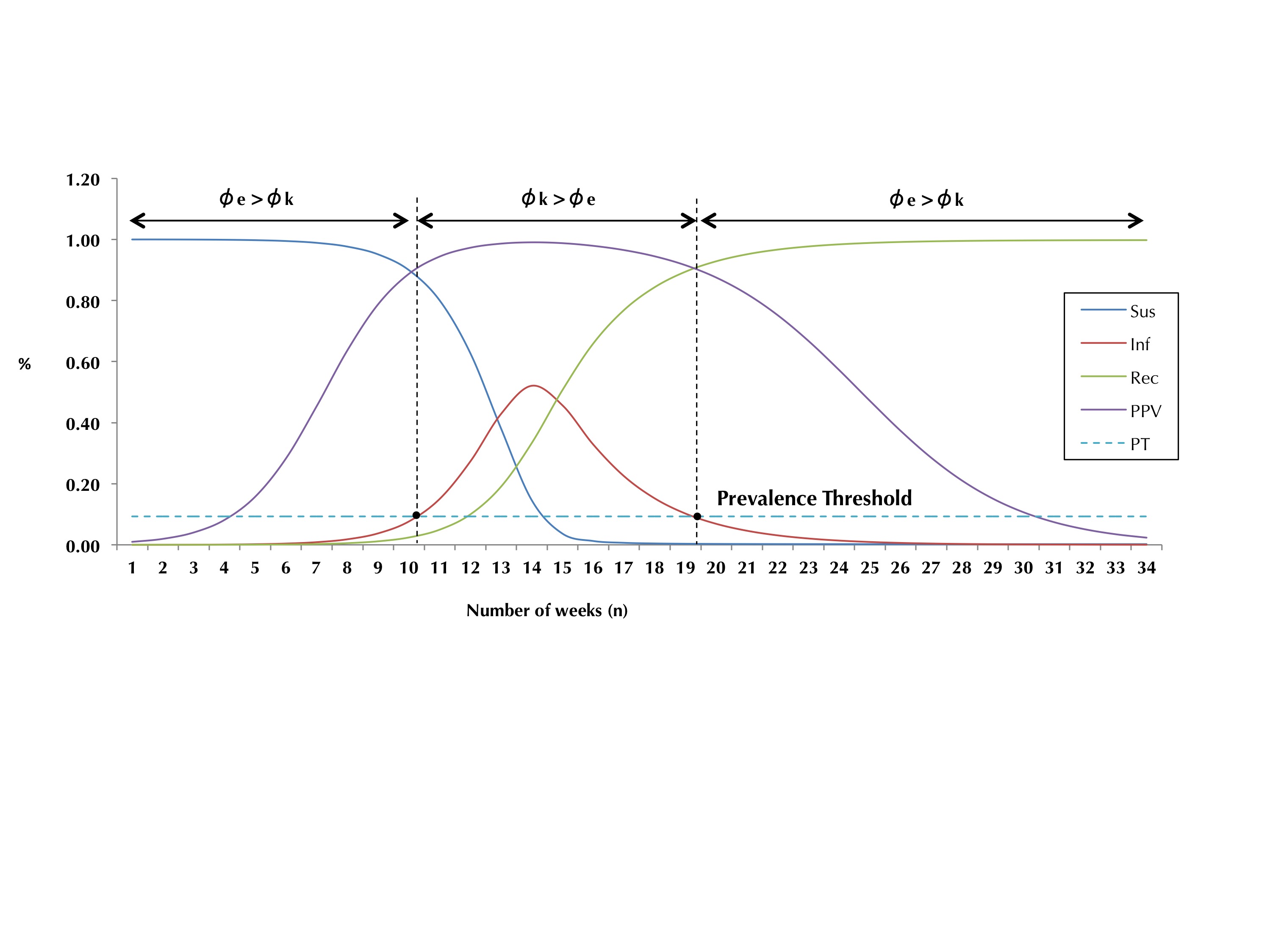}
\end{figure}
\end{landscape}
\newpage

Note the rise of the PPV (purple) as a function of prevalence (red). The delineation of the prevalence threshold (PT) at week 10 shows a corresponding flattening of the PPV, which holds steady almost horizontally. We thus observe that as the prevalence crosses the PT, the test performs well, with greater than 90 percent predictive value.  However, once the prevalence drops below the PT once again, around week 19, the PPV begins to drop anew. Of note, this is a consequence of the success of the screening test in the first place - leading to the accurate detection of disease in a higher proportion of individuals once the prevalence threshold has been crossed and people being adequately treated or quarantined to prevent further spread. In other words - as stated before -  by performing and succeeding at the very task it was developed to do, a screening test paradoxically reduces its predictive ability to correctly identify individuals with the disease it screens for in the future. The degree to which this paradoxical effect is observed depends on where we set our original prevalence, $\phi_0$, since $\zeta(\phi)$ depends on both $\phi_0$ and $k$. 
\\
\

Likewise, we can use the SIR model to come demonstrate the dynamics of the epidemic of disease X in real-time, numerically, as observed in Table 1. The disease manifests over a number of weeks, affecting a peak 52 percent of this population by week 14. Because implicit to the paradox is the fact that $\phi_0>\phi_k$, let us for argument's sake take the maximum prevalence as the starting prevalence point $\phi_0$ - though this need not be necessarily the case since the principles described in this work apply regardless of $\phi_0$. The time $k$ thus corresponds to $\phi_k$ each subsequent week. The corresponding PPV and the ensuing $\zeta(\phi_0,k)$ values can be seen in Table 1. Note that since our test has a sensitivity of 0.95 and a specificity of 0.99, Youden's $J$ statistic equals 0.95+0.99-1 = 0.94. Finally, we take the ceiling function n iteration number to ensure that we obtain an integer number of positive test iterations (PTI) needed to surpass the prevalence threshold - thus enhancing the reliability of the screening process \cite{balayla2020derivation}. Note that at the extremes of prevalence we would need to obtain 3 serial positive tests to achieve a PPV similar to that beyond the prevalence threshold. Once that threshold has been crossed, by definition $n_{i\phi_e}$ = 1. As noted above, other than developing newer, better screening tests, serial testing is one way to overcome the screening paradox \cite{balayla2020derivation} - be it with the same test done repeatedly or using a second, different diagnostic test altogether. 
\\
\

\newpage
\begin{center}
\textbf{Table 1. Numerical representation of the SIR model}
\end{center}
\begin{table}[ht!]
\centering
\begin{tabular}{cccccccc}
\hline
\multicolumn{1}{|c|}{\textbf{Week ($t$)}} & \multicolumn{1}{c|}{\textbf{PT}} & \multicolumn{1}{c|}{\textbf{Susceptible}} & \multicolumn{1}{c|}{\textbf{Infected}} & \multicolumn{1}{c|}{\textbf{Recovered}} & \multicolumn{1}{c|}{\textbf{PPV}} & \multicolumn{1}{c|}{\textbf{$\zeta(\phi_0,k)$}} & \multicolumn{1}{c|}{\textbf{$n_{i\phi_e}$}} \\ \hline
\multicolumn{1}{|c|}{1}             & \multicolumn{1}{c|}{0.093}       & \multicolumn{1}{c|}{0.9999}               & \multicolumn{1}{c|}{0.0001}            & \multicolumn{1}{c|}{0.0000}             & \multicolumn{1}{c|}{0.0094}       & \multicolumn{1}{c|}{0.0095}        & \multicolumn{1}{c|}{3.00}       \\ \hline
\multicolumn{1}{|c|}{2}             & \multicolumn{1}{c|}{0.093}       & \multicolumn{1}{c|}{0.9998}               & \multicolumn{1}{c|}{0.0002}            & \multicolumn{1}{c|}{0.0000}             & \multicolumn{1}{c|}{0.0196}       & \multicolumn{1}{c|}{0.0198}        & \multicolumn{1}{c|}{3.00}       \\ \hline
\multicolumn{1}{|c|}{3}             & \multicolumn{1}{c|}{0.093}       & \multicolumn{1}{c|}{0.9995}               & \multicolumn{1}{c|}{0.0004}            & \multicolumn{1}{c|}{0.0001}             & \multicolumn{1}{c|}{0.0405}       & \multicolumn{1}{c|}{0.0409}        & \multicolumn{1}{c|}{3.00}       \\ \hline
\multicolumn{1}{|c|}{4}             & \multicolumn{1}{c|}{0.093}       & \multicolumn{1}{c|}{0.9988}               & \multicolumn{1}{c|}{0.0009}            & \multicolumn{1}{c|}{0.0003}             & \multicolumn{1}{c|}{0.0816}       & \multicolumn{1}{c|}{0.0824}        & \multicolumn{1}{c|}{3.00}       \\ \hline
\multicolumn{1}{|c|}{5}             & \multicolumn{1}{c|}{0.093}       & \multicolumn{1}{c|}{0.9975}               & \multicolumn{1}{c|}{0.0020}            & \multicolumn{1}{c|}{0.0006}             & \multicolumn{1}{c|}{0.1577}       & \multicolumn{1}{c|}{0.1592}        & \multicolumn{1}{c|}{2.00}       \\ \hline
\multicolumn{1}{|c|}{6}             & \multicolumn{1}{c|}{0.093}       & \multicolumn{1}{c|}{0.9946}               & \multicolumn{1}{c|}{0.0041}            & \multicolumn{1}{c|}{0.0012}             & \multicolumn{1}{c|}{0.2829}       & \multicolumn{1}{c|}{0.2857}        & \multicolumn{1}{c|}{2.00}       \\ \hline
\multicolumn{1}{|c|}{7}             & \multicolumn{1}{c|}{0.093}       & \multicolumn{1}{c|}{0.9887}               & \multicolumn{1}{c|}{0.0087}            & \multicolumn{1}{c|}{0.0026}             & \multicolumn{1}{c|}{0.4541}       & \multicolumn{1}{c|}{0.4585}        & \multicolumn{1}{c|}{2.00}       \\ \hline
\multicolumn{1}{|c|}{8}             & \multicolumn{1}{c|}{0.093}       & \multicolumn{1}{c|}{0.9764}               & \multicolumn{1}{c|}{0.0181}            & \multicolumn{1}{c|}{0.0055}             & \multicolumn{1}{c|}{0.6372}       & \multicolumn{1}{c|}{0.6433}        & \multicolumn{1}{c|}{2.00}       \\ \hline
\multicolumn{1}{|c|}{9}             & \multicolumn{1}{c|}{0.093}       & \multicolumn{1}{c|}{0.9508}               & \multicolumn{1}{c|}{0.0376}            & \multicolumn{1}{c|}{0.0115}             & \multicolumn{1}{c|}{0.7878}       & \multicolumn{1}{c|}{0.7955}        & \multicolumn{1}{c|}{2.00}       \\ \hline
\multicolumn{1}{|c|}{10}            & \multicolumn{1}{c|}{0.093}       & \multicolumn{1}{c|}{0.8993}               & \multicolumn{1}{c|}{0.0766}            & \multicolumn{1}{c|}{0.0241}             & \multicolumn{1}{c|}{0.8874}       & \multicolumn{1}{c|}{0.8960}        & \multicolumn{1}{c|}{2.00}       \\ \hline
\multicolumn{1}{|c|}{11}            & \multicolumn{1}{c|}{0.093}       & \multicolumn{1}{c|}{0.8002}               & \multicolumn{1}{c|}{0.1502}            & \multicolumn{1}{c|}{0.0496}             & \multicolumn{1}{c|}{0.9438}       & \multicolumn{1}{c|}{0.9529}        & \multicolumn{1}{c|}{1.00}       \\ \hline
\multicolumn{1}{|c|}{12}            & \multicolumn{1}{c|}{0.093}       & \multicolumn{1}{c|}{0.6271}               & \multicolumn{1}{c|}{0.2733}            & \multicolumn{1}{c|}{0.0997}             & \multicolumn{1}{c|}{0.9728}       & \multicolumn{1}{c|}{0.9822}        & \multicolumn{1}{c|}{1.00}       \\ \hline
\multicolumn{1}{|c|}{13}            & \multicolumn{1}{c|}{0.093}       & \multicolumn{1}{c|}{0.3803}               & \multicolumn{1}{c|}{0.4289}            & \multicolumn{1}{c|}{0.1908}             & \multicolumn{1}{c|}{0.9862}       & \multicolumn{1}{c|}{0.9957}        & \multicolumn{1}{c|}{1.00}       \\ \hline
\multicolumn{1}{|c|}{14}            & \multicolumn{1}{c|}{0.093}       & \multicolumn{1}{c|}{0.1454}               & \multicolumn{1}{c|}{0.5208}            & \multicolumn{1}{c|}{0.3337}             & \multicolumn{1}{c|}{0.9904}       & \multicolumn{1}{c|}{1.0000}        & \multicolumn{1}{c|}{1.00}       \\ \hline
\multicolumn{1}{|c|}{15}            & \multicolumn{1}{c|}{0.093}       & \multicolumn{1}{c|}{0.0364}               & \multicolumn{1}{c|}{0.4563}            & \multicolumn{1}{c|}{0.5074}             & \multicolumn{1}{c|}{0.9876}       & \multicolumn{1}{c|}{0.9972}        & \multicolumn{1}{c|}{1.00}       \\ \hline
\multicolumn{1}{|c|}{16}            & \multicolumn{1}{c|}{0.093}       & \multicolumn{1}{c|}{0.0125}               & \multicolumn{1}{c|}{0.3281}            & \multicolumn{1}{c|}{0.6594}             & \multicolumn{1}{c|}{0.9789}       & \multicolumn{1}{c|}{0.9884}        & \multicolumn{1}{c|}{1.00}       \\ \hline
\multicolumn{1}{|c|}{17}            & \multicolumn{1}{c|}{0.093}       & \multicolumn{1}{c|}{0.0066}               & \multicolumn{1}{c|}{0.2246}            & \multicolumn{1}{c|}{0.7688}             & \multicolumn{1}{c|}{0.9649}       & \multicolumn{1}{c|}{0.9743}        & \multicolumn{1}{c|}{1.00}       \\ \hline
\multicolumn{1}{|c|}{18}            & \multicolumn{1}{c|}{0.093}       & \multicolumn{1}{c|}{0.0044}               & \multicolumn{1}{c|}{0.1519}            & \multicolumn{1}{c|}{0.8437}             & \multicolumn{1}{c|}{0.9445}       & \multicolumn{1}{c|}{0.9536}        & \multicolumn{1}{c|}{1.00}       \\ \hline
\multicolumn{1}{|c|}{19}            & \multicolumn{1}{c|}{0.093}       & \multicolumn{1}{c|}{0.0035}               & \multicolumn{1}{c|}{0.1022}            & \multicolumn{1}{c|}{0.8943}             & \multicolumn{1}{c|}{0.9154}       & \multicolumn{1}{c|}{0.9242}        & \multicolumn{1}{c|}{1.00}       \\ \hline
\multicolumn{1}{|c|}{20}            & \multicolumn{1}{c|}{0.093}       & \multicolumn{1}{c|}{0.0030}               & \multicolumn{1}{c|}{0.0687}            & \multicolumn{1}{c|}{0.9284}             & \multicolumn{1}{c|}{0.8751}       & \multicolumn{1}{c|}{0.8835}        & \multicolumn{1}{c|}{2.00}       \\ \hline
\multicolumn{1}{|c|}{21}            & \multicolumn{1}{c|}{0.093}       & \multicolumn{1}{c|}{0.0027}               & \multicolumn{1}{c|}{0.0461}            & \multicolumn{1}{c|}{0.9513}             & \multicolumn{1}{c|}{0.8210}       & \multicolumn{1}{c|}{0.8290}        & \multicolumn{1}{c|}{2.00}       \\ \hline
\multicolumn{1}{|c|}{22}            & \multicolumn{1}{c|}{0.093}       & \multicolumn{1}{c|}{0.0025}               & \multicolumn{1}{c|}{0.0309}            & \multicolumn{1}{c|}{0.9666}             & \multicolumn{1}{c|}{0.7517}       & \multicolumn{1}{c|}{0.7590}        & \multicolumn{1}{c|}{2.00}       \\ \hline
\multicolumn{1}{|c|}{23}            & \multicolumn{1}{c|}{0.093}       & \multicolumn{1}{c|}{0.0024}               & \multicolumn{1}{c|}{0.0207}            & \multicolumn{1}{c|}{0.9769}             & \multicolumn{1}{c|}{0.6676}       & \multicolumn{1}{c|}{0.6741}        & \multicolumn{1}{c|}{2.00}       \\ \hline
\multicolumn{1}{|c|}{24}            & \multicolumn{1}{c|}{0.093}       & \multicolumn{1}{c|}{0.0023}               & \multicolumn{1}{c|}{0.0139}            & \multicolumn{1}{c|}{0.9838}             & \multicolumn{1}{c|}{0.5720}       & \multicolumn{1}{c|}{0.5775}        & \multicolumn{1}{c|}{2.00}       \\ \hline
\multicolumn{1}{|c|}{25}            & \multicolumn{1}{c|}{0.093}       & \multicolumn{1}{c|}{0.0023}               & \multicolumn{1}{c|}{0.0093}            & \multicolumn{1}{c|}{0.9884}             & \multicolumn{1}{c|}{0.4713}       & \multicolumn{1}{c|}{0.4758}        & \multicolumn{1}{c|}{2.00}       \\ \hline
\multicolumn{1}{|c|}{26}            & \multicolumn{1}{c|}{0.093}       & \multicolumn{1}{c|}{0.0022}               & \multicolumn{1}{c|}{0.0062}            & \multicolumn{1}{c|}{0.9915}             & \multicolumn{1}{c|}{0.3731}       & \multicolumn{1}{c|}{0.3768}        & \multicolumn{1}{c|}{2.00}       \\ \hline
\multicolumn{1}{|c|}{27}            & \multicolumn{1}{c|}{0.093}       & \multicolumn{1}{c|}{0.0022}               & \multicolumn{1}{c|}{0.0042}            & \multicolumn{1}{c|}{0.9936}             & \multicolumn{1}{c|}{0.2847}       & \multicolumn{1}{c|}{0.2874}        & \multicolumn{1}{c|}{2.00}       \\ \hline
\multicolumn{1}{|c|}{28}            & \multicolumn{1}{c|}{0.093}       & \multicolumn{1}{c|}{0.0022}               & \multicolumn{1}{c|}{0.0028}            & \multicolumn{1}{c|}{0.9950}             & \multicolumn{1}{c|}{0.2102}       & \multicolumn{1}{c|}{0.2123}        & \multicolumn{1}{c|}{2.00}       \\ \hline
\multicolumn{1}{|c|}{29}            & \multicolumn{1}{c|}{0.093}       & \multicolumn{1}{c|}{0.0022}               & \multicolumn{1}{c|}{0.0019}            & \multicolumn{1}{c|}{0.9959}             & \multicolumn{1}{c|}{0.1512}       & \multicolumn{1}{c|}{0.1527}        & \multicolumn{1}{c|}{2.00}       \\ \hline
\multicolumn{1}{|c|}{30}            & \multicolumn{1}{c|}{0.093}       & \multicolumn{1}{c|}{0.0022}               & \multicolumn{1}{c|}{0.0013}            & \multicolumn{1}{c|}{0.9966}             & \multicolumn{1}{c|}{0.1065}       & \multicolumn{1}{c|}{0.1076}        & \multicolumn{1}{c|}{2.00}       \\ \hline
\multicolumn{1}{|c|}{31}            & \multicolumn{1}{c|}{0.093}       & \multicolumn{1}{c|}{0.0022}               & \multicolumn{1}{c|}{0.0008}            & \multicolumn{1}{c|}{0.9970}             & \multicolumn{1}{c|}{0.0739}       & \multicolumn{1}{c|}{0.0747}        & \multicolumn{1}{c|}{3.00}       \\ \hline
\multicolumn{1}{|c|}{32}            & \multicolumn{1}{c|}{0.093}       & \multicolumn{1}{c|}{0.0022}               & \multicolumn{1}{c|}{0.0006}            & \multicolumn{1}{c|}{0.9973}             & \multicolumn{1}{c|}{0.0508}       & \multicolumn{1}{c|}{0.0512}        & \multicolumn{1}{c|}{3.00}       \\ \hline
\multicolumn{1}{|c|}{33}            & \multicolumn{1}{c|}{0.093}       & \multicolumn{1}{c|}{0.0022}               & \multicolumn{1}{c|}{0.0004}            & \multicolumn{1}{c|}{0.9974}             & \multicolumn{1}{c|}{0.0346}       & \multicolumn{1}{c|}{0.0349}        & \multicolumn{1}{c|}{3.00}       \\ \hline
\multicolumn{1}{|c|}{34}            & \multicolumn{1}{c|}{0.093}       & \multicolumn{1}{c|}{0.0022}               & \multicolumn{1}{c|}{0.0003}            & \multicolumn{1}{c|}{0.9976}             & \multicolumn{1}{c|}{0.0234}       & \multicolumn{1}{c|}{0.0236}        & \multicolumn{1}{c|}{3.00}       \\ \hline
\multicolumn{1}{l}{}                & \multicolumn{1}{l}{}             & \multicolumn{1}{l}{}                      & \multicolumn{1}{l}{}                   & \multicolumn{1}{l}{}                    & \multicolumn{1}{l}{}              & \multicolumn{1}{l}{}               & \multicolumn{1}{l}{}            \\
\multicolumn{8}{l}{Model for a screening test with 95 $\%$ sensitivity and 99 $\%$ specificity over time $t$.}                                                                                                                                                                                                   \\
\multicolumn{8}{c}{PPV = positive predictive value, $\zeta(\phi)$ is the PPV ratio between $\phi_0$}                                                                                                                                                                                                                  \\
\multicolumn{8}{c}{and $\phi_k$, N = number of iterations to overcome the screening paradox}                                                                                                                                                                                                                    
\end{tabular}
\end{table}
\newpage
\section{Conclusion}
In this manuscript, we explore the mathematical model which formalizes the screening paradox and explore its implications for population level screening programs as a function of the position of the initial prevalence of a condition  relative to the prevalence threshold level of its screening test. Likewise, we provide a mathematical model to determine the predictive value percentage loss as the prevalence decreases and define the number of positive test iterations (PTI) needed to reverse the effects of the paradox when a single test is undertaken serially. Given their theoretical nature, clinical application of the concepts herein reported need validation prior to implementation. Meanwhile, an understanding of how the dynamics of prevalence can affect the PPV over time can help inform clinicians as to the reliability of a screening test's results.
\newpage

\bibliographystyle{unsrt}
\bibliography{references}
\end{document}